\documentclass[a4paper,12pt]{article}
\usepackage{cite}
\newcommand{\be}{\begin{equation}}
\newcommand{\ee}{\end{equation}}
\newcommand{\bea}{\begin{eqnarray}}

\newcommand{\eea}{\end{eqnarray}}

\begin{document}

\title{
\begin{flushright}
{\small UB-ECM-PF-04/33}\\
\end{flushright}\vspace{1.5cm}
When conceptual worlds collide: \\ The GUP and the BH entropy}

\author{A.J.M. Medved$\;^a$ and
Elias C. Vagenas$\;^b$ \\ \\
$\;^a$~School of Mathematics, Statistics and Computer Science \\
Victoria University of Wellington \\
PO Box 600, Wellington, New Zealand \\
E-Mail: joey.medved@mcs.vuw.ac.nz \\ \\
$\;^b$~Departament d'Estructura i Constituents de la Mat\`{e}ria\\
and\\ CER for Astrophysics, Particle Physics and Cosmology\\
Universitat de Barcelona\\
Av. Diagonal 647, E-08028 Barcelona, Spain\\
E-Mail: evagenas@ecm.ub.es\\ \\}

\maketitle
\begin{abstract}

Recently, there has been much  attention devoted to resolving
the quantum corrections
to the  Bekenstein--Hawking (black hole) entropy.
In particular, many researchers have expressed a vested
interest in fixing the coefficient  of  the sub-leading
logarithmic term. In the current paper, we are able to make
some substantial progress in this  direction by utilizing
the generalized uncertainty principle (GUP).
Notably, the GUP reduces to the conventional
Heisenberg relation in situations of weak gravity but
transcends it  when gravitational effects can no longer be
ignored. Ultimately, we  formulate  the
quantum-corrected entropy in terms of an expansion
that is consistent with all previous findings. Moreover, we demonstrate
that  the logarithmic prefactor (indeed, any
coefficient of the expansion) can be expressed in terms of a single parameter
that  should be determinable via the fundamental theory.

\end{abstract}
\section{Introduction}

One of the  most remarkable achievements in gravitational physics
was  the realization that black holes are thermodynamic
objects with a well-defined entropy and temperature \cite{Bek,Bek2,Haw}.
As our current interest is with the entropy,
let us recall the famous Bekenstein--Hawking formula:
\be
S_{BH}={A\over 4 L_p^2}\;.
\label{bhe}
\ee
To be perfectly clear, $A$ represents the
cross-sectional area of the black hole horizon, $L_p=\sqrt{\hbar G}$
is the Planck
length, while the speed of light   and Boltzmann's constant
are always set to unity. Also, we will always assume, for the sake of
clarity, a macroscopically
large Schwarzschild black hole in a four-dimensional spacetime.

It is worth noting that the main arguments in support of
equation (\ref{bhe})
are purely of a thermodynamic (rather than statistical) nature.
But let us suppose, as has become common in the literature,
that the Bekenstein--Hawking entropy  can be attributed a definite statistical
meaning.~\footnote{However, for some commentary on
why this might not be the case, see \cite{Brief}.}
Then how might one go about identifying these microstates and, even more
optimistically, counting them? The answer to this question presumably lies
within the framework of the elusive fundamental theory;
also known as quantum gravity.  Indeed, the two leading candidate
theories --- namely, string theory  and loop quantum gravity
---  have both
had  success (albeit, with caveats attached) at statistically explaining
the entropy--area ``law'' ({\it e.g.}, respectively, \cite{StV,ASh}).

The proponents of either of these theories proclaim this success at
state counting  to
be one of the  major achievements of their favored program.
However, what about the unbias observer, who might actually  prefer
if there was only one fundamental theory? In this regard, one might
be inclined to call upon the quantum corrections to  $S_{BH}$ (which, in spite
of its intrinsically quantum origins, is a tree-level quantity \cite{GM}).
On general grounds --- and has been verified in a multitude of
studies~\footnote{See \cite{Ref} for an extensive list of references.}
---  one would expect that the quantum-corrected entropy takes on the
following expansive form:
\be
S =\frac{A}{4L_{p}^{2}}+c_{0}\ln
\left(\frac{A}{4L_{p}^{2}}\right)
+\sum_{n=1}^{\infty}c_{n} \left({A\over 4 L_p^2}
\right)^{-n}+ \rm{const}\;,
\ee
where the coefficients $\{c_n\}$ can be regarded as model-dependent
parameters.
Most of the recent focus has been on $c_0$; that is, the  coefficient
of the leading-order
correction or the logarithmic ``prefactor''. It has even been suggested that
this particular parameter might be useful as  a discriminator of
prospective fundamental theories \cite{KM}. It is, therefore, appropriate
to reflect upon  the loop-quantum-gravity  prediction of $c_o= -1/2$
(according to the most up-to-date rigorous calculation
\cite{Mer}); whereas string theory makes no
similar type of assertion that we are aware of.~\footnote{Note that the
loop-quantum-gravity result refers strictly to the microcanonical
correction. There will also be a canonical correction,
irrespective of the fundamental theory, that contributes
at least $+1/2$ to $c_0$ \cite{Kas,GMe,CM}.}
However, without any further input, how can we say if
any particular value  of the prefactor (such as $c_0=-1/2$)
is right or wrong? That is, unlike the tree-level calculations,
this type of discrimination is based on asking a question
for which we do not yet know the answer!

It becomes clear that, to proceed in this direction, one requires
a method of fixing $c_0$ (as well as the lower-order
coefficients) that does not depend on the specific elements
of any one particular model of quantum gravity.~\footnote{Let us
point out  two recent studies that have endeavored
to at least put restrictions on the value of the logarithmic prefactor
\cite{XXX,Hod}. Nevertheless, the former was specific to
loop quantum gravity whereas the latter used a premise
that is believed to be contradictory with the same theory.
Hence, the general utility of their results are subject
to question.} For instance, it might be hoped that the holographic principle
\cite{Bou} could serve just such a purpose.  In the current paper, we will
utilize a similarly general ({\it i.e.}, model-independent)  concept;
namely, the {\it generalized uncertainty principle} or the GUP
(see \cite{Gar} for a general discussion).
The premise of the GUP (which will be presented in due course) is
that, as gravity is turned on,
the ``conventional''  Heisenberg relation is no longer
completely satisfactory (but  still perfectly valid,
in an approximate sense, in low-gravity regimes).  Although the GUP
has its  historical origins in string theory \cite{String1,String2}
(as well as non-commutative quantum mechanics
\cite{ncqm1,ncqm2}),
it can also be argued for on the basis of simple {\it Gedanken} experiments
that make  no reference to the specifics of the fundamental theory
\cite{Ged1,Ged2}.
This means that the GUP is conceptually ideal for realizing the
discussed objective.~\footnote{Further
citations on the GUP can be found in
\cite{CD}. The reader might also find \cite{CA,AAP,setare}  as being helpful
precursors to the upcoming analysis.}

The remainder of the paper goes as follows. Using the GUP
as our primary input, we present
a perturbative calculation of the quantum-corrected entropy,
which can  readily be extended to any desired order.
Note that, here,  we  consider strictly the microcanonical
corrections,~\footnote{For an earlier discussion on interpreting black hole
thermodynamics in a mircocanonical framework, see \cite{HY}.
For an example of calculating
quantum corrections to the entropy from a substantially different
perspective, see \cite{SV}.}
as the canonical corrections have  been dealt with
elsewhere
({\it e.g.}, \cite{Kas,GMe,CM,More}).
The paper concludes with
a discussion that emphasizes the relevance of our outcome.

\section{Analysis}

Let us begin the formal analysis by presenting the GUP as it
typically  appears in the literature ({\it e.g.}, \cite{Gar}); namely,
\be
\delta x\geq {\hbar\over \delta p}+\alpha L_{p}^2 {\delta p\over \hbar}\;.
\label{gup}
\ee
Here, $\delta x$ and $\delta p$ are the position and momentum uncertainty
for a quantum  particle,
and $\alpha$ is a dimensionless (probably model-dependent) constant of
the order unity.
Let us point out that the combination $L_{p}^2/\hbar$ can be replaced
with the Newtonian coupling  constant $G$,
thus implying that the ``extra'' (right-most) term
is truly a consequence of gravity.  Which is to say,
the presumption of a gravitational modification
to the quantum uncertainty principle --- along with dimensional considerations
--- is sufficient to fix the {\it form} of equation (\ref{gup});
with the parameter $\alpha$ reflecting our remaining ignorance.
[It should be noted that, on general grounds, an infinite series of
higher-order corrections
can be anticipated on the right-hand side of equation (\ref{gup}).
However, as explained in the footnote at the end of Section 2,
this caveat does not affect our conclusions.]

By way of
some simple manipulations,  we can re-express the GUP
in the following manner (subsequently setting $\hbar=1$):
\be
\delta p \geq \frac{\delta x}{2\alpha^{2}L_{p}^{2}}
\left[1-\sqrt{1-\frac{4\alpha^{2}L_{p}^{2}}{\left(\delta x\right)^{2}}}\right]
\hspace{1ex},
\label{gup1}
\ee
where a sign choice has been made by imposing the correct classical
($L_p\rightarrow 0$) limit.
Since $L_p$ is  normally  viewed as an ultraviolet cutoff on the
spacetime geometry ({\it e.g.},\cite{LY}),
it should be safe to  regard the dimensionless ratio $L_{p}^2/(\delta x)^2$
as small relative to unity.
[This is certainly true for our regime of interest --- see equation (\ref{YYY})
below.]  Hence, it is natural to Taylor expand the square root and obtain
\be
\delta p \geq \frac{1}{\delta x}
\left[1+\left(\frac{\alpha^{2}L_{p}^{2}}{\left(\delta x\right)^{2}}\right)+
2\left(\frac{\alpha^{2}L_{p}^{2}}
{\left(\delta x\right)^{2}}\right)^{2}+\cdots\right]
\label{gup2}
\hspace{1ex}.
\ee

Next, let us consider the following measurement process: a photon is used to
ascertain the position
of a quantum particle of energy $E$.
Starting with this  setup, one can call upon a standard textbook
argument \cite{LL}
(also see \cite{AAP})  to translate
the ``conventional''  uncertainty principle (or $\delta p\geq 1/\delta x$)
into the  lower bound $E\geq 1/\delta x\;$.
Now generalizing to the case of strong gravity and thus invoking
the GUP, we have
\be
E\geq  \frac{1}{\delta x}
\left[1+\left(\frac{\alpha^{2}L_{p}^{2}}{\left(\delta x\right)^{2}}\right)+
2\left(\frac{\alpha^{2}L_{p}^{2}}
{\left(\delta x\right)^{2}}\right)^{2}+\cdots\right]
\label{gup3}
\hspace{1ex}.
\ee

Let us now apply the above formalism to a specific case
which is particularly relevant to our current interest.
More to the point, we will consider a quantum particle that starts out
in the vicinity
of an event horizon  and then is ultimately absorbed by the black hole.
But let us first take note of the general-relativistic result
that, for a  black hole absorbing a classical particle of
energy $E$ and size $R$, the minimal increase in
the horizon area  can be expressed as \cite{Chris}
\be
\left(\Delta A\right)_{min} \geq 8\pi  L_p^2 E R
\label{minarea1}
\hspace{1ex}.
\ee
Given such a classical context,
one is of course free to  set $R=0$, and so  this
is really no bound at all. Nonetheless, as originally
observed by Bekenstein \cite{Bek2},
there is no such freedom for a quantum particle since
$R$ can never be taken as smaller than $\delta x$ ({\it i.e.}, the
intrinsic uncertainty in the position of the particle).
Hence, for the case of a quantum particle, one obtains the {\it finite} bound
\be
\left(\Delta A\right)_{min} \geq 8\pi  L_p^2 E \delta x
\label{minarea11}
\hspace{1ex}.
\ee

Substituting equation (\ref{gup3}) into the above,
we then  have, as a consequence of the GUP,
\be
\left(\Delta A\right)_{min} \geq 8\pi L_{p}^{2}
\left[1+
\left(\frac{\alpha^{2}L_{p}^{2}}{\left(\delta x\right)^{2}}\right)+
2 \left(\frac{\alpha^{2}
L_{p}^{2}}{\left(\delta x\right)^{2}}\right)^{2}+\cdots
\right]
\label{minarea2}
\ee
or
\be
\left(\Delta A\right)_{min} \simeq \epsilon L_{p}^{2}
\left[1+
\left(\frac{\alpha^{2}L_{p}^{2}}{\left(\delta x\right)^{2}}\right)+
2 \left(\frac{\alpha^{2}
L_{p}^{2}}{\left(\delta x\right)^{2}}\right)^{2}+\cdots
\right] \;.
\label{minarea22}
\ee
In the latter form, $\epsilon$ is a yet-to-be-determined  numerical factor
that is greater than (but of the order of) $8\pi$.
Note that, on statistical grounds, however, one might well argue
for $\epsilon=4\ln(k)\;$  such that $k$ is  a strictly positive integer
\cite{Muk}.

The pressing concern is now what to take for $\delta x$.
We can best address this issue by first  reclarifying our objective:
it is to deduce the entropy of a (macroscopically large) Schwarzschild
black hole
of {\it fixed} mass; that is, compute the {\it microcanonical} entropy.
Such a framework necessitates a black hole that is (by some means)
emersed in a bath of radiation at precisely its
own temperature.
(Otherwise, there would be a net gain or loss of mass
with time, and a  microcanonical framework would no longer be appropriate.)
Hence, the particles that we are interested in have a
Compton length on the order of
the inverse of the Hawking temperature \cite{Haw} (as measured
by an asymptotic observer, which is implicit in calculations of this nature).
Actually, the inverse surface gravity ($\kappa^{-1}= 2r_{Sch}$
where $r_{Sch}$ is the Schwarzschild radius) is probably the most
sensible choice of length scale in the context of near-horizon geometry.
On this basis, let us choose
\be
\delta x \sim 2  r_{Sch}\;.
\label{YYY}
\ee
(Note that an uncertainty of this order has been previously argued for
on different  but probably related grounds; {\it e.g.},
\cite{AD,KF}.)
Granted, there is some degree of ambiguity here; nevertheless, as
discussed below,
this is not much of a concern.

Realizing that $(\delta x)^2 \sim  A/\pi$, we can rewrite equation
({\ref{minarea22}) as follows:
\be
\left(\Delta A\right)_{min}
\simeq\epsilon L_{p}^{2}\left[1+
\left(\frac{\pi\alpha^{2}L_{p}^{2}}{A}\right)+
2\left(\frac{\pi\alpha^{2}L_{p}^{2}}{A}\right)^{2}+\cdots\right]
\;.
\label{minarea3}
\ee
Admittedly,
we have used an input [equation (\ref{YYY})] that is, quite possibly,
off by an order-of-unity numerical factor.
But let us take note of the form of the expansion in equation
({\ref{minarea22}); in particular,  we always have the
ratio $\alpha^2/(\delta x)^2\;$.
Hence,
it should be  clear  that any such numerical
discrepancy in $\delta x$
can be systematically ``absorbed'' (without loss of generality) into the
already ambiguous parameter $\alpha$.
Which is to say, we can still regard  equation (\ref{minarea3})
as an accurate statement (up to the explicit order)
modulo  the intrinsic uncertainty in  the parameters $\alpha$
and  $\epsilon$.

In accordance with the ideas of information theory ({\it e.g.}, \cite{Adami}),
one would anticipate that the minimal increase of entropy should be,
irrespective of the value of the area, simply  one ``bit'' of
information; let us denote this fundamental unit of
entropy  as $b$. (Typically, $b=\ln 2$,
but we need not be precise on this point.)
Then, inasmuch as the black hole entropy should depend strictly on the
horizon area \cite{Bek2}, it follows that
\be
\frac{d S}{d A}
\simeq
\frac{\left(\Delta S\right)_{min}}{\left(\Delta A\right)_{min}}
\simeq
\frac{b}
{\epsilon L_{p}^{2}\left[1+
\left(\frac{\pi\alpha^{2}L_{p}^{2}}{A}\right)+
2\left(\frac{\pi\alpha^{2}L_{p}^{2}}{A}\right)^{2}+\cdots\right]}
\hspace{1ex}
\label{ratio1}
\ee
or
\be
\frac{d S}{d A} \simeq
\frac{b}{\epsilon L_{p}^{2}}
\left[1-\left(\frac{\pi\alpha^{2}
L_{p}^{2}}{A}\right)-\left(\frac{\pi\alpha^{2}
L_{p}^{2}}{A}\right)^{2}+\cdots\right] \;.
\label{ratio2}
\ee

Now integrating the above, we have (up to a constant term)
\be
S\simeq \frac{A}{4L_{p}^{2}} - {\pi\alpha^2\over 4}\ln \left({A\over 4  L_p^2}
\right)
+\left({\pi\alpha^2\over 4}\right)^2
\left(\frac{4 L_{p}^{2}}{A}\right)
+\cdots
\hspace{1ex},
\label{entropy}
\ee
where the Bekenstein--Hawking area law {\cite{Bek,Bek2,Haw} has been used to
calibrate
$b/\epsilon =1/4$. Note that one can also write
\be
S =\frac{A}{4L_{p}^{2}}-{\pi\alpha^{2}\over 4}\ln
\left(\frac{A}{4L_{p}^{2}}\right)
+\sum_{n=1}^{\infty}c_{n} \left({A\over 4 L_p^2}
\right)^{-n}+ \rm{const}\;,
\label{last}
\ee
where  the expansions coefficients $c_n\propto \alpha^{2(n+1)}$ can always be
computed to any desired order
of accuracy.~\footnote{Let us now remind the reader that
the GUP should, itself, probably be regarded as a power-law expansion.
For instance, there is an alternative formulation of equation
(\ref{gup}) that goes as  $\delta x\delta p \geq \sqrt{1+2\alpha
L_{p}^2(\delta p)^2}\simeq 1+\alpha L_p^2 (\delta p)^2-
{1\over 2} \alpha^2L_p^4(\delta p)^4+\cdots$ \cite{ncqm1}.
In this event, the calculation
of the numerical factors in the coefficients of  equation (\ref{last})
would be technically more complicated than previously suggested.
Nonetheless, such a calculation could certainly  be
carried out and our main observation,   $c_n\propto \alpha^{2(n+1)}$,
would remain intact. Moreover, the logarithmic-order coefficient
is completely unaffected by such considerations.}

\section{Conclusion}

In summary, we have utilized the generalized uncertainty principle (or the GUP)
to demonstrate an explicit form for the {\it quantum-corrected}
black hole entropy.
Let us now make some
pertinent comments about our result: \\

{\it (i)} We have obtained a leading-order correction to the classical
entropy--area law that goes as the logarithm of the area. This is consistent
with numerous other studies that have delved into this subject matter.
(Again, consult \cite{Ref} for a list of references.)
\\

{\it (ii)} Unlike many other treatments, we have achieved
a concise algorithm for calculating  the sub-leading (or inverse power-law)
corrections to the entropy. Moreover, this calculation can be carried out to
any perturbative order; with the expansions coefficients depending on
only a single parameter --- namely, $\alpha^ 2$.
\\

{\it(iii)} Let us re-emphasize that our entropic
calculation was a {\it microcanonical} one.
This  is consistent with
the logarithmic correction being negative; that is, the microcanonical
framework necessitates some type of additional boundary conditions
(or gauge fixing), which naturally implies a reduction in the entropy.
Moreover, a process of quantization (as was implicit in our use
of a quantum uncertainty relation) can be expected, on general grounds,
to  remove entropy
from the system. On the other hand, any {\it canonical} corrections to
the entropy would most certainly be positive; as this class
can be attributed  to thermal fluctuations in the horizon area.
(See \cite{Gour,CMNew,Log}
for further elaboration on these points.)
If one were to (somehow) measure the entropy of a ``real'' black hole,
it is likely that both types of corrections would have to be accounted for.
Nonetheless, it is probably fair to say that the microcanonical class
is essentially the more fundamental one. \\

{\it(iv)} A controversial issue in the literature is
the exact value of the logarithmic ``prefactor'' ({\it i.e.},
the coefficient of the $\ln A$ term).  Here, we have found
that it takes on the value  $-\pi\alpha^{2}/4$ (as had earlier been deduced
in a related but distinct treatment \cite{CD,CDM});
where $\alpha$
is an order-of-unity parameter that reflects our ignorance about
the exact form of the GUP ---  which is to say, our ignorance
about the underlying  theory of quantum gravity.
In spite of the current lack of knowledge about $\alpha$,
our calculation could still have merit as a discriminator of
prospective theories of quantum gravity. To elaborate,
the fundamental theory should  (at least in principle) be able to
make a precise statement about $\alpha$ and,  as a consequence of our
calculation,
a prediction about the logarithmic-order coefficient.   Moreover, the theory of
quantum gravity
should also have something to say about this coefficient
through more direct means; that is, through a process of state counting.
The success (or failure) of these two calculations to match up
could then be viewed as an important consistency check (or a revealing
conceptual flaw) for any testable candidate theory.

\newpage

\section*{Acknowledgments}

Research for AJMM is supported  by
the Marsden Fund (c/o the  New Zealand Royal Society)
and by the University Research  Fund (c/o Victoria University).
ECV would like to thank Professors Saurya Das, G. Amelino-Camelia,
R. Adler, and P. Chen for  useful correspondences and enlightening comments.
Both authors thank Professor M. Cavaglia for making an important
observation.

\end{document}